\documentclass[useAMS,usenatbib]{mn2e}
\usepackage{graphicx}

\begin{document}


\title[A large sample of LSB disk galaxies]
{A large sample of low surface brightness disk galaxies from the
SDSS. I: The sample and the stellar populations}

\author[Zhong et al.]{G. H. Zhong,$^{1,2}$\thanks{E-mail: ghzhong@bao.ac.cn}
Y. C. Liang,$^{1}$\thanks{E-mail: ycliang@bao.ac.cn} F. S.
Liu,$^{4,1}$ F. Hammer,$^{3}$ J. Y. Hu,$^{1}$ X. Y. Chen,$^{1}$
\newauthor L. C. Deng$^{1}$ and B. Zhang$^{2,1}$
\\
$^1$National Astronomical Observatories, Chinese Academy of
Sciences,20A Datun Road, Chaoyang District, Beijing 100012, China\\
$^2$Department of Physicals,Hebei Normal University,Shijiazhuang 050016, China \\
$^3$GEPI, Observatoire de Paris-Meudon, 92195 Meudon, France\\
$^4$College of Physics Science and Technology, Shenyang Normal
University, Shenyang 110034, China }

\date{Accepted 2008 Sep.17 Received2008 Sep. 16}

\pagerange{\pageref{firstpage}--\pageref{lastpage}} \pubyear{2008}

\maketitle

\label{firstpage}

\begin{abstract}
We present the properties of a large sample (12,282) of nearly
face-on low surface brightness (LSB) disk galaxies selected from
the main galaxy sample of SDSS-DR4. These properties include
$B$-band central surface brightness $\mu_0(B)$, scale lengths $h$,
integrated magnitudes, colors, and distances $D$. This sample has
$\mu_0(B)$ values from 22 to 24.5 mag arcsec$^{-2}$ with a median
value of 22.42 mag arcsec$^{-2}$, and disk scale lengths
ranging from 2 to 19 kpc. They are quite bright with $M_B$ taking
values from -18 to -23 mag with a median value of -20.08 mag.
There exist clear correlations between log$h$ and $M_B$, log$h$
and log$D$, log$D$ and $M_B$.  However, no obvious correlations
are found between $\mu_0(B)$ and log$h$, colors etc. The correlation
between colors and log$h$ is weak even though it exists.
Both the optical-optical and optical-NIR color-color diagrams
indicate that most of them have a mixture of young and old stellar
populations. They also satisfy color-magnitude relations, which
indicate that brighter galaxies tend generally to be redder. The
comparison between the LSBGs and a control sample of nearly
face-on disk galaxies with higher surface brightness (HSB) with
$\mu_0(B)$ from 18.5 to 22 mag arcsec$^{-2}$ show that, at a given
luminosity or distance, the observed LSB galaxies tend to have
larger scale lengths. These trends could be seen gradually by
dividing both the LSBGs and HSBGs into two sub-groups according to
surface brightness.
A volume-limited sub-sample was extracted to check the
incompleteness of surface brightness. The only one of the property
relations having an obvious change is the relation of log$h$
versus $\mu_0(B)$, which shows a correlation in this sub-sample.
\end{abstract}

\begin{keywords}
galaxies: distances and redshifts -
          galaxies: fundamental parameters -
          galaxies: photometry -
          galaxies: spiral -
          galaxies: stellar content
\end{keywords}

\section{Introduction}

Low surface brightness galaxies (LSBGs) are galaxies that emit
much less light per area than normal galaxies. Yet, owing to their
faintness compared with the night sky, they are hard to find.
Hence their contribution to the local galaxy population  has been
underestimated for a long time. It has been suggested that LSBGs
may comprise up to half of the local galaxy population (McGaugh et
al. 1995).

The discovery and searching history of the LSBGs and the
understanding on them have been summarized by Impey $\&$ Bothun
(1997) and Bothun et al.(1997). The first quantitative suggestion
about LSBGs is the so-called Freeman's law. It was noticed by
Freeman (1970) that the central surface brightness of 28 out of 36
disk galaxies fell in the range of $\mu_0(B)$=21.65$\pm$0.3
mag~arcsec$^{-2}$. This could be caused by selection effects
(Disney 1976; Zwicky 1957).

The successful hunting for very diffuse galaxies (Romanishin et
al. 1983 (from UGC); Sandage et al. 1984; Ellis et al. 1984; Bothun et al.
1987) prompted some new surveys. The first one was based on the
second Palomar Sky Survey (POSS-II, Schombert and Bothun 1988;
Schombert et al. 1992). A second one was initiated in the
Fornax cluster using the Malinization technique in order to
compare the results to Virgo (Bothun et al. 1992; Caldwell $\&$
Bothun 1987). The third one used the Automated Plate Measuring
(APM) to scan UK Schmidt plates using an algorithm optimized to
find LSB galaxies. This formed the most extensive catalog of LSB
galaxies to that date (Impey et al. 1996). Morshidi-Esslinger et
al. (1999a,b) extended such sample. The ``Texas survey'' (O'Neil
et al. 1997a,b) increases the surface density of LSBGs in the
general field, in which they firstly found the red LSB galaxy
populations.

Following these catalogs, various aspects of LSBGs have been
investigated. Especially, in a series of publications, Impey and
his colleagues analyzed their APM survey sample (693 field
galaxies in various morphologies, see Impey et al. 1996;
Sprayberry et al. 1996; Sprayberry et al. 1997; Impey et al. 2001;
Burkholder et al. 2001). Numerous other studies about LSBGs have
been done based on smaller samples, relating to surface photometry
and color distributions (McGaugh $\&$ Bothun 1994; de Blok et al.
1995; O'Neil et al. 1997a);
 number, luminosity and mass density
(McGaugh 1996); AGN properties (Mei et al. 2008); and in the
optical (Beijersbergen et al. 1999; Bergvall et al. 1999; Galaz et
al. 2006, Zackrisson et al. 2005, Pizzella et al. 2008), near-IR (NIR, Bergvall et al.
1999; Galaz et al. 2002,2006; Monnier Ragaigne et al. 2003,
Zackrisson et al. 2005), infrared (Hinz et al. 2007) and UV (Boissier et al. 2008)
wavelengths.  In addition, Kniazev et al. (2004) developed a
method to search for LSBGs from modern survey and used the sample
of Impey et. al. (1996) to test their method.

These studies show that LSBGs may be unevolved systems with low
metallicity (McGaugh 1994; de Blok $\&$ van der Hulst 1998a;
O'Neil et al. 1998; Bell et al. 2000), low stellar formation rate
(van der Hulst et al. 1993; van Zee et al. 1997), small stellar
density, a relatively high gas fraction (de Blok et al. 1996;
McGaugh $\&$ de Blok 1997) and large amounts of dark matter (de
Blok $\&$ McGaugh 1997). Despite these impressive progresses,
there are still several challenges about LSBGs, such as many
aspects of their formation and evolution, in particular,
considerable uncertainty regarding their star formation history
and so on. Moreover, the previous studies were based on small
samples, up to about several hundreds, while the present modern
digital sky surveys, such as the Sloan Digital Sky Survey (SDSS),
greatly improved the observed numbers of astronomical objects, and
could provide a much larger sample of LSBGs, hence giving much
more information. Therefore, we propose a project to search for a
large sample of LSBGs from the SDSS database, and then to study
their properties in detail.

In this first paper of our series about the large sample of LSBGs
selected from the SDSS, we will present how to select the sample
and then show their basic properties and stellar populations from
photometric colors. Then, we will present their spectroscopic
properties (Liang et al. 2008, Chen et al. 2008, in preparation) 
and their detailed
surface photometry and color gradients (Zhong et al. 2008, in
preparation) etc. This paper is organized as follows. In
Sect.~\ref{sec.2}, we describe how to select the sample. In
Sect.3, we present the relations between some property parameters.
In Sect.4, we compare the LSBGs with
 the HSB disk galaxies sample selected
simultaneously. In Sect.5, we discuss the incompleteness of the
sample at the low surface brightness end and present the property
relations of a volume-limited sub-sample. In Sect.6, we do cross
correlation for the sample with the 2MASS database to obtain their
optical and NIR color-color diagrams and the color-magnitude
relations. The discussions and conclusions are given in Sect.7.
Throughout the paper, a cosmological model with $H_0$=70 km
s$^{-1}$ Mpc$^{-1}$, $\Omega_M$=0.3  and $\Omega_\lambda$=0.7 is
adopted. The units of surface brightness is mag~arcsec$^{-2}$.

\section[]{The Sample}
\label{sec.2}

The SDSS is the most ambitious astronomical survey ever undertaken
in imaging and spectroscopy (York et al. 2000; Stoughton et al.
2002; Abazajian et al. 2003, 2004). The imaging data are done in
drift scan mode and are 95\% complete for point sources at 22.0,
22.2, 22.2, 21.3, and 20.5 in five bands ($u$, $g$, $r$, $i$ and
$z$) respectively. The spectra are flux- and wavelength-calibrated
with 4096 pixels from 3800 to 9200 \AA~at $R\approx1800$.

The sample used in this work is selected from the main galaxy
sample (MGS) of SDSS-DR4 (Strauss et al. 2002), which comprises
galaxies with  $r$-band Petrosian magnitude $r\leq$17.77
(corrected for foreground Galactic extinction using the reddening
maps of Schlegel et al. 1998) and $r$-band Petrosian half-light
surface brightness $\mu_{50}\leq$24.5 mag arcsec$^{-2}$. This
sample has a median redshift of about 0.10.

The MGS is a spectroscopic sample selected from the SDSS
photometric data. The completeness for objects with spectroscopy
observations is high, exceeding 99\%, and the fraction of galaxies
eliminated by surface brightness cut is very small ($\sim0.1\%$).
The only significant source of incompleteness identified is
bending with saturated stars, which is higher for higher
brightness because they subtend more sky (Strauss et al. 2002).
Relative to all the SDSS targets, the SDSS spectroscopic survey is
90\% complete (Blanton et al. 2003; Hogg et al. 2004; Strauss et
al. 2002; McIntosh et al. 2006). The 7\% incompleteness coming
primarily from galaxies missed due to fiber collisions since the
minimum separation of fiber centers is 55$\arcsec$ (Blanton et al.
2003). It does lead to a slight underrepresentation of
high-density regions (Hogg et al. 2004), however, where bright
early-type galaxies usually dominate. Additionally, there are a
small faction of redshift failures ($\sim1\%$) and some bright
star contamination ($\sim2\%$) that becomes more significant for
brighter galaxies (Strauss et al. 2002; McIntosh et al. 2006).
However, the completeness of the sample at the low surface
brightness end needs to be discussed carefully because of the
known problem of surface brightness limit in a redshift survey and
the focus of our work on the low surface brightness ones. We will
specially discuss this incompleteness in Sect.~\ref{sec.55} and
will extract a volume-limited sub-sample, which is complete, from
our whole sample galaxies to check what changes appearing in the
related properties of the galaxies.

We prefer to use the DR4 rather than the latest DR6 here. That is
because DR4 has been large enough for statistical analysis of
LSBGs; other property parameters for galaxies in DR4 have also
been obtained by the MPA/JHU
group\footnote{http://www.mpa-garching.mpg.de/SDSS/}, such as the
emission-line measurements etc.; and the
NYU-VAGC\footnote{http://sdss.physics.nyu.edu/vagc/} catalog
(Blanton et al. 2005) has made some useful cross correlations with
other survey databases. These can provide us more information on
both photometry and spectroscopy. Our sample galaxies are selected
as follows.

\subsection[]{The whole sample: the nearly face-on disk galaxies}
\label{sec.2.1}

To obtain reliable surface brightness values of the disk galaxies
and minimize the effect of dust extinction inside the galaxies, we
select the nearly face-on disk galaxies as a working sample by
following the criteria below.

\begin{enumerate}

\item  $fracDev_r$ $<$ 0.25

The parameter $fracDev_r$ indicates the fraction of luminosity
contributed by the de Vaucouleurs profile relative to exponential
profile in the $r$-band. The topic in this work focuses on disk
galaxies, whose surface brightness profiles usually can be well
described by an exponential formula (e.g., Bernardi et al. 2005;
Chang et al. 2006b; Shao et al. 2007). Therefore, we select the
sample galaxies almost having an exponential light profile by
requiring their $fracDev_r$ $<$ 0.25. This can minimize the effect
of bulge light on the disk galaxies. Using this selection criterion,
we can obtain 111,479 ($\sim$28\%) objects from the total 401,007
galaxies in the SDSS-DR4 main galaxy sample.

\item $b/a$ $>$ 0.75

To select nearly face-on galaxies, we also require objects with
the axis ratio $b/a$$>$0.75 (corresponding to the inclination
$i<$41.41 degree), where $a$ and $b$ are the semi-major and
semi-minor axes of the fitted exponential disk respectively. After
this step is applied, 32,226 ($\sim$29\% of 111,479) galaxies are
left.

\item $M_B$ $<$ -18

Keeping the $B$-band absolute magnitude $M_B$ $<$ -18 in mind
excludes the few dwarf galaxies contained in the sample. The
$B$-band absolute magnitude $M_B$ is calculated using
\begin{equation}
\label{eq.Mb} M = m - 5\log (D_L) - K(z),
\end{equation}
where $m$ is the apparent magnitude that corrected for Galactic
extinction using the reddening maps of Schlegel et al. (1998),
$D_L$ is the luminosity distance calculated by using the Local
Group relative redshift values from the NYU-VAGC catalog (Blanton
et al. 2005, see Sect.~\ref{sec.3.1}). The $K$-correction $K$($z$)
are calculated by using the $K$-corrections program given on the
NYU-VAGC website (Blanton et al. 2005). Finally, we obtain
30,333 galaxies as our whole sample after excluding $\sim$6\% of
32,226 nearly face-on disk galaxies by this criterion.

\end{enumerate}

\subsection[]{The subsamples: the LSBGs and HSBGs}
\label{sec.2.2}

A commonly used parameter to classify a galaxy to be one with low
(or high) surface brightness is its $B-$band central surface
brightness $\mu_0(B)$. In this work, we adopted $\mu_0(B)$$\geq$22
mag arcsec$^{-2}$, a commonly applied criterion (Boissier et al.
2003), as LSBGs and $\mu_0(B)$$<$22 mag arcsec$^{-2}$ as high
surface brightness galaxies (HSBGs) following Impey, Burkholder \&
Sprayberry (2001) and O'Neil et al. (1997a,b).

The surface brightness profiles of disk galaxies are well
approximated by an exponential disk profile with the form:
\begin{equation}
\Sigma(r) = \Sigma_0 exp(-{{r}\over{a}}),
\end{equation}
where $\Sigma_0$ is the surface brightness of the disk in units of
$M_\odot$ pc$^{-2}$ and $a$ is the disk scale-length measured in
units of arcsec. In the logarithmic units, this equation becomes:
\begin{equation}
\mu_r = \mu_0 + 1.086({{r}\over{a}}),
\end{equation}
where $\mu_0$ is the central surface brightness in mag
arcsec$^{-2}$.

In the analysis, the disk is assumed to be infinitely thin. The
total flux is given by:
\begin{equation}
F_{tot} = 2 {\pi} a^2 \Sigma_0.
\end{equation}
Then, the total apparent magnitude can be determined from this
equation. Converting this to a logarithmic scale is:
\begin{equation}
\mu_0 = m + 2.5\log {(2 \pi a^2)},
\end{equation}
where $m$ refers to the total apparent magnitude and $\mu_0$ refers
to the central surface brightness.

The central surface brightness is corrected by inclination and
cosmological dimming effects by following:
\begin{equation}
\mu_0 = m + 2.5\log {(2 \pi a^2)} + 2.5\log{(b/a)} - 10\log{(1+z)},
\end{equation}
where $z$ is the redshift of the target.

\begin{figure}
\centering
\includegraphics [width=7.5cm, height=5.2cm] {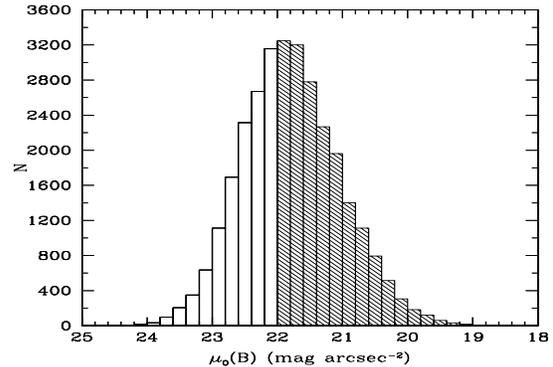}
\caption{Histogram distribution of the selected nearly face-on
disk galaxies (30,333) in $B$-band central surface brightness
$\mu_0(B)$. The whole sample is divided into two parts: the white
 region with $\mu_0(B)$ $\geq$22 mag arcsec$^{-2}$ as the
LSBGs (12,282), and the shadowed region with $\mu_0(B)$ $<$22 mag
arcsec$^{-2}$ as the HSBGs (18,051). } \label{fig.1}
\end{figure}

The $B$-band (also $U$, $V$, $R$ ) central surface brightnesses
and magnitudes of the SDSS galaxies can be calculated from the $g$
and $r$-band quantities by using the conversions provided by Smith
et al. (2002). The histogram distribution of $\mu_0(B)$ for our
selected whole sample (30,333 galaxies) is shown in
Fig.~\ref{fig.1}. The blank histogram is for the LSBGs and the
shadowed one is for the HSBGs with the separating value of 22 mag
arcsec$^{-2}$. We call the LSBGs as {\bf Sample-L} and the HSBGs
as {\bf Sample-H} in this work. Sample-L includes 12,282
($\sim$40\%) galaxies with 22$\leq$$\mu_0(B)$$\leq$24.5 mag
arcsec$^{-2}$ with median and mean values of 22.42 and 22.50 mag
arcsec$^{-2}$ respectively. Sample-H includes 18,051 ($\sim$60\%)
galaxies with 18.5$\leq$$\mu_0(B)$$<$22 mag arcsec$^{-2}$ with
median and mean values of 21.41 and 21.29 mag arcsec$^{-2}$
respectively. Since the LSBGs are the major topic of this paper,
we will carefully present the properties of Sample-L in
Sect.~\ref{sec.3} and roughly compare the HSBGs with the LSBGs in
Sect.~\ref{sec.4}. However, when looking for the relations of some
property parameters with surface brightness, it is useful to show
all of the sample, including LSBGs and HSBGs, therefore, in
Sect.3.2 we present both the LSBGs and HSBGs in the relations of
$m_B$, distance, $M_B$ with $\mu_0$(B). The cross-correlations
with near infrared 2MASS data will be emphasized in
Sect.~\ref{sec.5}.

\section[]{Property parameters and relations of the large sample of LSBGs}
\label{sec.3}

Some property parameters of LSBGs have been obtained and discussed
in previous work, such as Impey et al. (1996, 2001), de Blok et
al. (1995), Bothun et al. (1997), Galaz et al. (2002, 2006) etc.
 In this section, we present some property parameters and
relations for our Sample-L (the LSBGs) and compare them with the
previous results. The Sample-H (HSBGs) are also given in the
relations of $m_B$, distance, $M_B$ with $\mu_0$(B) (see Sect.3.2
and Fig.~\ref{fig.3}), which is useful to show the entire sample.

\subsection[]{The histogram distributions of some property parameters}
\label{sec.3.1}

\begin{figure}
\centering
\includegraphics [width=7.8cm, height=7.5cm] {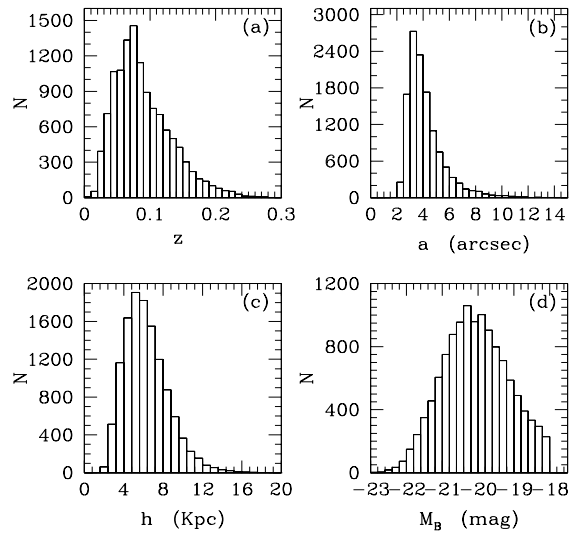}
\caption{Histogram
distributions of some property parameters of Sample-L (the LSBGs):
(a). the redshift, obtained from NYU-VAGC catalog and is the
peculiar velocity-corrected Local Group relative redshift from
SDSS; (b). the disk scale length (the disk semi-major axis) in
arcsec as $a$; (c). the disk scale length (the radius) in kpc as
$h$; (d). the $B$-band absolute magnitude $M_B$, which obviously
shows that LSBGs does not necessarily have low luminosity, and the
LSBGs can be very luminous.} 
\label{fig.2}
\end{figure}

We show the distributions of some property parameters of our LSBGs
in Fig.~\ref{fig.2}. Fig.~\ref{fig.2}a shows the histogram
distribution of their redshift $z$ obtained from the NYU-VAGC
catalog (Blanton et al. 2005), which are the peculiar velocity
corrected Local Group relative redshifts from SDSS. It shows that
the redshifts are from 0 to 0.3 with median and mean values of
0.080 and 0.089, respectively.
Fig.~\ref{fig.2}b shows the histogram distributions of the disk
scale-length $a$ in arcsec, and Fig.~\ref{fig.2}c shows the
histogram distributions of the disk scale-length $h$ in kpc. The
range of $a$ is from 2 to 13 arcsec with median and mean values of
3.78 and 4.19 arcsec, and $h$ is from 2 kpc to 19 kpc with median
and mean values of 5.98 and 6.30 kpc, respectively. The scale
lengths of LSBGs investigated here are in the same ranges as in
McGaugh (1992) and de Blok et al. (1995). Fig.~\ref{fig.2}d shows
the histogram distribution of the $B$-band absolute magnitudes,
which is calculated using Eq.~(\ref{eq.Mb}). It shows that the
LSBGs are not necessary faint galaxies, but rather they are quite
bright, $M_B$ is from -18 to -23 mag with the median and mean
values of -20.08 and -20.06 mag, respectively.

\subsection{Distributions of some property parameters with $B$-band central surface brightness}

We present the relations between the total $B$-band apparent
magnitudes, distance, absolute magnitude $M_B$ and the $B$-band
central surface brightness $\mu_0(B)$ for Sample-L and
Sample-H.

\subsubsection{central surface brightness and apparent magnitude}

Figure~\ref{fig.3}a presents the relations between the $B$-band
apparent magnitudes and the $B$-band central surface brightness
$\mu_0(B)$. The vertical lines separate the LSBGs and HSBGs with
$\mu_0(B)$=22 mag arcsec$^{-2}$. It shows that these galaxies have
the $m_B$ from 14 to 19 mag. At a given magnitude, most galaxies
are detectable over a wide range of surface brightnesses, for
example, 18$<$$\mu_0(B)$$<$24.5 mag arcsec$^{-2}$ at $m_{B}$$=$18
mag. There exists almost no correlation within the ranges. Some
objects missed probably at faint $\mu_0$ could be due to the poor
contrast with the sky brightness. These results are similar to
those shown by the samples of APM survey (693 LSB field galaxies)
given by Impey et al. (1996).

\subsubsection{central surface brightness and distance}

Figure~\ref{fig.3}b presents the distribution of the distance
versus $\mu_0(B)$ of the sample galaxies. For the LSBGs, the left
part of the vertical line, it shows that the range of surface
brightness for the nearest galaxies is substantially larger than
that for the more distant galaxies. We expect this to influence
surface brightness selection, where galaxies with higher surface
brightness could be seen at the largest distances (Disney $\&$
Phillips 1983, Impey et al. 1996). But for the HSBGs, the right
part of the vertical line, it does not obviously show such trend.
This may mean that the selection bias affects much the LSBGs.
 This
distribution is consistent with that of the sample in the APM
survey presented by Impey et al. (1996). However, the
volume-limited sub-sample of LSBGs do not show such a trend (see
Sect.5).

\subsubsection{central surface brightness and absolute magnitude}

\begin{figure}
\centering
\includegraphics[bb= 56 128 252 728, width=6.0cm, height=11.2cm, clip] {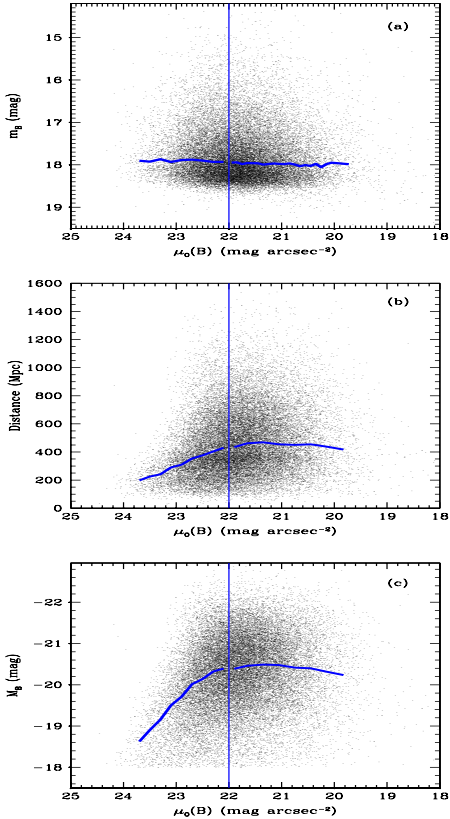}
\caption{Distributions of some property parameters with $\mu_0(B)$
for our LSBGs (the left part of the vertical line) and HSBGs (the
right part of the vertical line): (a). the total $B$-band apparent
magnitudes; (b). the distance of the galaxies; (c). the $B$-band
absolute magnitudes. The solid lines are the median-value points
within the bins of 0.10 of $\mu_0(B)$ less than 23.6 mag
arcsec$^{-2}$, and we take $\mu_0(B)$ greater than 23.6 mag
arcsec$^{-2}$ as one bin because there are very few sources.
The lines refer to the median-value points within the bins. }
 \label{fig.3}
\end{figure}

Figure~\ref{fig.3}c presents the distribution of the absolute
$B$-band magnitudes versus $\mu_0(B)$ of the sample galaxies. The
left part from the vertical line is for the LSBGs. There exists a
correlation among them showing that the fainter galaxies tend to
have lower surface brightnesses though there is large scatter. It
also shows over the whole range of 22$<$$\mu_0(B)$$<$24.5 mag
arcsec$^{-2}$, galaxies are being discovered over the entire range
of -23$<$$M_B$$<$-18 mag. The right part from the vertical line is
for the HSBGs, which show that there is no obvious correlation
between their $M_B$ and $\mu_0(B)$. This could mean that the
selection bias affects the LSBGs greatly. These are also similar
to that presented by the APM sample given by Impey et al. (1996).

\subsection{Distributions of some property parameters with disk scale length and distance}

Some of the previous studies have shown there are obvious
correlations between the luminosity and disk scale length log$h$
(Impey et al. 2001; Bergvall et al. 1999), distance log$D$
and disk scale length log$h$ (Impey et al. 2001) for LSBGs.
These relations become more clear from our much larger and
homogeneous sample.

\subsubsection{disk scale length and luminosity}
\label{sec.3.3.1}
\begin{figure}
\centering
\includegraphics [width=8.0cm, height=5.8cm] {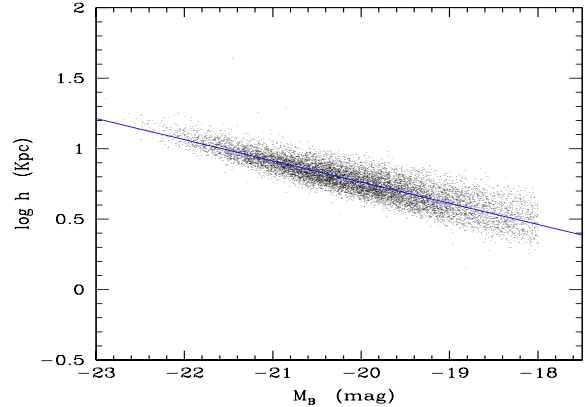}
\caption{Relations of the LSB sample galaxies about their disk
scale length in logarithm versus $B$-band absolute magnitudes. The
solid line refers to the least-square fit for the sample galaxies
given by Eq.(\ref{eqhMb}). The Spearman-Rank Order
correlation coefficient is -0.893.}
\label{fig.4}
\end{figure}

We show the relation between absolute magnitudes $M_B$ and disk
scale length log$h$ for our Sample-L in Fig.~\ref{fig.4}. It can
be seen that there is a very clear correlation between them, which
can be fitted by a least-square fit as
\begin{eqnarray}
 \log h = -0.150(\pm 0.0007) M_B - 2.245(\pm 0.014),
\label{eqhMb}
\end{eqnarray}
with a small standard deviation of 0.070 dex. The
Spearman-Rank Order correlation coefficient is -0.893.

This correlation has been shown in Impey et al. (2001, for their
238 galaxies from APM) and Bergvall et al. (1999, for the gathered
samples from literature), but is more obvious for our larger and
homogeneous sample. It shows brighter galaxies tend to have
larger scale lengths.

It is worth comparing such relations of our LSBGs with the
corresponding relations of HSBGs. They show some discrepancies,
which will be specially discussed in Sect.~\ref{sec.4}.

\subsubsection{disk scale length and distance}

\begin{figure}
\centering
\includegraphics [width=8.0cm, height=5.8cm] {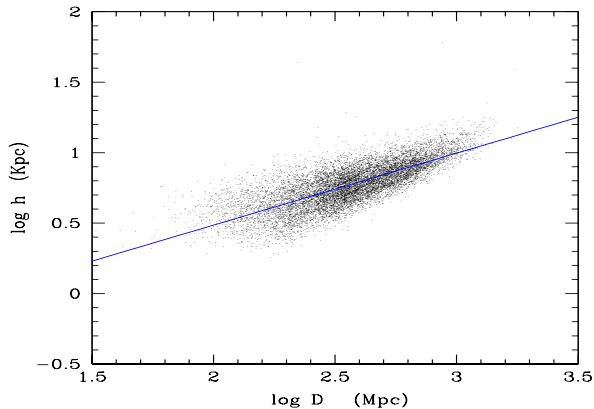}
\caption{Relations of the LSB sample galaxies about their distance
versus disk scale length. The solid line refers to the
least-square fit for the sample galaxies given by Eq.(\ref{eqhD}).
The Spearman-Rank Order correlation coefficient is 0.767.}
\label{fig.5}
\end{figure}

Our LSBGs sample also show a very clear correlation between their
distance and disk scale length, which is given in
Fig.~\ref{fig.5}. This correlation can be given by  a least-square
fit as:
\begin{eqnarray}
\log h = 0.511(\pm 0.004) \log D - 0.536(\pm 0.010),
\label{eqhD}
\end{eqnarray}
with the standard derivation of 0.10 dex. The Spearman-Rank
Order correlation coefficient is 0.767.

This relation obviously show the selection effect, i.e., in the
more distant Universe, the galaxies with larger scale lengths have
the benefit that they can be detected and observed.

\subsubsection{distance and luminosity}

\begin{figure}
\centering
\includegraphics [width=8.0cm, height=5.8cm] {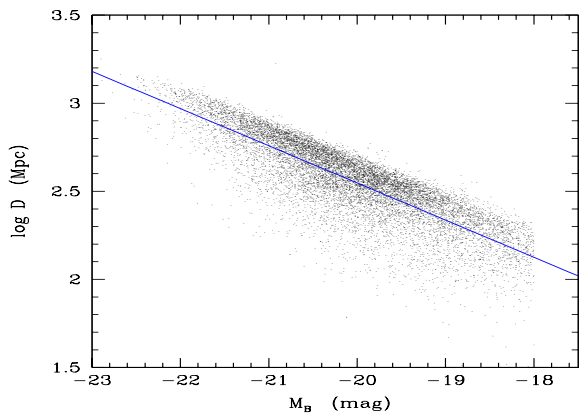}
\caption{Distribution of the distance log$D$ and the $B$-band
absolute magnitude $M_B$ for our LSBGs. The solid line refers to
the least-square fit for the sample galaxies given by
Eq.(\ref{eq.DMb}). The Spearman-Rank Order correlation
coefficient is -0.847.}
\label{fig.13}
\end{figure}

Figure~\ref{fig.13} shows the distribution between the distance
log$D$ and the $B$-band absolute magnitude $M_B$. It could be seen
that there is an obvious correlation between them, though there is
some scatter. This correlation can be shown by a least-square fit
as:
\begin{eqnarray}
\log D = -0.211(\pm 0.001) M_B - 1.675(\pm 0.025),
\label{eq.DMb}
\end{eqnarray}
with the standard derivation of 0.13 dex. The Spearman-Rank
Order correlation coefficient is -0.847.

This correlation show the selection effects of the survey
observation. That means the brighter galaxies can be observed at
larger distances, and the relatively fainter ones can only be
observed at smaller distances.

\subsubsection{disk scale lengths and central surface brightness}

\begin{figure}
\centering
\includegraphics [width=8.0cm, height=5.8cm] {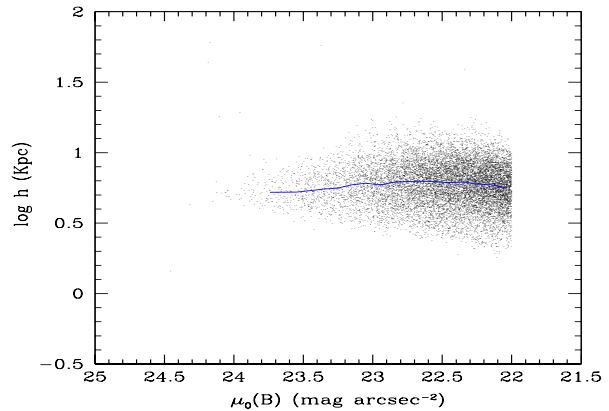}
\caption{Distribution of the disk scale length versus the $B$-band
surface brightness for our LSBGs. The solid line is the
median-value points within the bins that is the same as
Fig.~\ref{fig.3}.
      }
\label{fig.6}
\end{figure}

Figure~\ref{fig.6} shows the distribution between disk scale
length log$h$ and the central surface brightness $\mu_0(B)$, which
does not show obvious correlation. This result is consistent with
those of Beijerbergen et al. (1999) and O'Neil et al. (1997b) for
smaller samples.

\subsection{Relations between colors and disk scale length, colors and central surface brightness }

In this section we show the relations between colors and disk
scale length, as well as between colors and central surface
brightness for the LSBGs.

\subsubsection{colors and disk scale lengths }

\begin{figure}
\centering
\includegraphics [width=7.8cm, height=8.6cm] {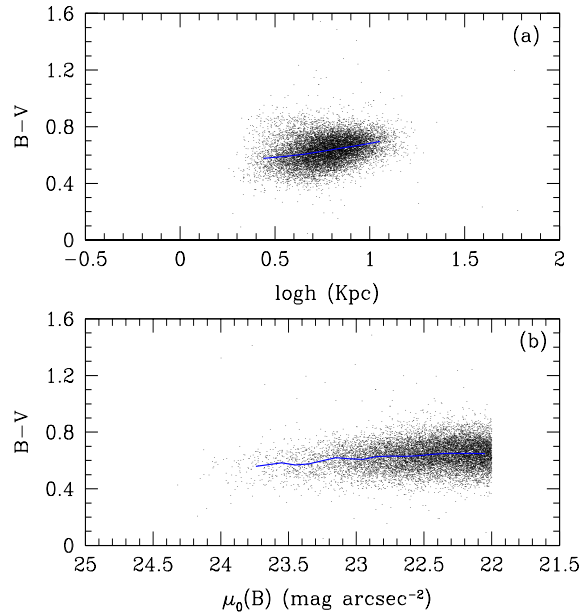}
\caption{(a). Distribution of $B - V$ color as a function of disk
scale length for the Sample-L. The solid line is the median-value
points within the bins of 0.10 of $\log h$ from 0.5 to 1.0 ($h$ in
kpc). But $\log h$ less than 0.5 or greater than 1.0 ($h$ in kpc)
as one bin, respectively, since the sources are very few in those
two bins. (b). Distribution of $B - V$ color as a function of
$B$-band central surface brightness for the Sample-L. The solid
line is the median-value points within the bins that is same as
Fig.~\ref{fig.3}. } 
\label{fig.7}
\end{figure}

 Figure~\ref{fig.7}a presents the distribution between $(B-V)$
color and disk scale length of our large sample of LSBGs. The
 correlation between them is weak even though it exists, and the
 Spearman-Rank Order correlation coefficient between them is only 0.302.
 Our result is not much different from the previous
studies on small samples (e.g., de Blok et al. 1995; Beijersbergen
et al. 1999), which showed the colors do not depend on sizes of
LSBGs. But Bergvall et al. (1999) have showed an obvious
correlation for a small sample of spirals, large LSBGs and their
blue LSBGs.

\subsubsection{colors and central surface brightness}

Figure~\ref{fig.7}b presents the distribution between the $(B-V)$
colors and central surface brightness $\mu_0(B)$ for our Sample-L,
which does not show obvious correlation in the whole ranges of the
two parameters. The results presented from smaller samples of
LSBGs studied by Beijersbergen et al. (1999), de Blok et al.
(1995) and Bothun et al. (1997) are consistent with our result
here from a large sample of LSBGs.

\subsection{The optical color-color diagram and stellar populations}

\subsubsection{The negligible affect of dust extinction on colors}
\label{sec.3.5.1}

\begin{figure}
\centering
\includegraphics [width=8.0cm, height=5.8cm] {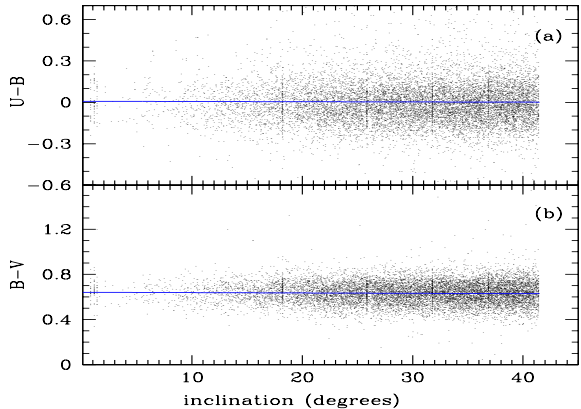}
\caption{ Relations between disk inclination and colors of
the Sample-L: (a). $U-B$, (b). $B-V$. The solid lines are the
least-square fits for the data, which show almost no slopes.
      }
\label{fig.12}
\end{figure}

In this work, we have selected the nearly face-on disk galaxies to
minimize the affection of dust extinction. Thus their properties
of stellar population won't be affected much by dust extinction,
which could be confirmed by the independence of the colors with
inclination. Fig.~\ref{fig.12} shows such independence of the LSB
sample galaxies (12,282). Indeed the HSB sample galaxies present
similar independence, which were not presented as plots here to
save space. Additionally, we also do cross-correlations for these
nearly face-on disk galaxies with the IRAS infrared PSCz catalog
within a range of 5 arcsec, but only find very few IR-detected
objects, which also confirm the negligible affect of dust on their
properties. This negligible affect of dust on colors is consistent
with other studies (e.g., Bell et al. 2000; Galaz et al. 2006).

\subsubsection{$UBV$ diagram and stellar populations}

O'Neil et al. (1997a,b) have performed a digital survey for LSBGs
in the spiral-rich Cancer and Pegasus clusters as well as the low
density regime defined by the Great Wall (the ``Texas survey").
They found a total of 127 galaxies of angular diameter larger than
15 arcseconds with $\mu_0(B)$$\geq$22 mag arcsec$^{-2}$. They
discussed the stellar populations of this sample of LSBGs by
optical color-color relations and firstly found the very red
population of LSBGs. They divided the galaxies into three
categories: the very blue (with $(U-B) < -0.2, (B-V) < 0.6$), the
very red (with $(U-B) > 0.3, (B-V) > 0.8$), and the rest that have
other color values.

\begin{figure}
\centering
\includegraphics [width=8cm, height=5.8cm] {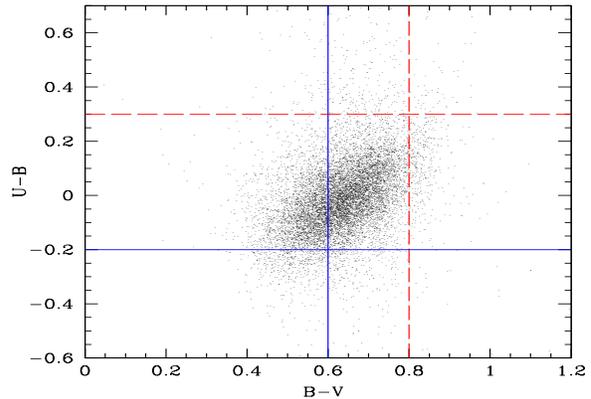}
\caption{$U-B$ versus $B-V$ diagram of our LSBGs. The solid and
dashed lines are taken from O'Neil et al. (1997b) to define the
``very blue" and ``very red" populations of galaxies (see text). }
\label{fig.8}
\end{figure}

In our Fig.~\ref{fig.8}, we shows the $(U-B)$ vs. $(B-V)$ diagram
for our Sample-L. The average values of our sample galaxies are
$<(U-B)> = -0.0012 \pm 0.08$, $<(B-V)> = 0.64 \pm 0.06$ in the
ranges of $-0.6 \leq (U-B) \leq 0.6$ and $0 \leq (B-V) \leq 1.2$.
These are not very different from O'Neil's sample, and is similar
to those of Romanishin et al. (1983) and McGaugh $\&$ Bothun
(1994) as well.

We also plot the separation lines suggested by O'Neil et al.
(1997a,b) for the ``very blue" and ``very red" LSBGs in
Fig.~\ref{fig.8} (the solid and dashed lines, see above). Our
Sample-L contains 535 ($\sim 4.4\%$) ``very blue" LSBGs and 69
($\sim 0.48\%$) ``very red" LSBGs. Most of the sample galaxies are
the ``rest case" with mixed stellar populations.

Heller $\&$ Brosch (2001) also present such $UBV$ colors for their
29 LSB dwarf irregular galaxies from the Virgo cluster, and found
their ``very blue" is about 30\%  and no ``very red" ones. These
two different fractions could be easily understood since their
samples are dwarf irregular galaxies, which should be blue. de
Blok et al. (1995) also present such a $UBV$ diagram for a small
sample of 21 LSB galaxies gathered from literature, and they did
not find any sample belonging to the ``very red" range that O'Neil
et al. suggested.

The colors of these LSBGs are obviously different from those of
the E0/S0 galaxies but not much different from the Sc or later
spiral galaxies with active star formation. As O'Neil et al.
(1997b) commented, the average colors of E0/S0 galaxies (good
examples of old stellar populations) have $(U-B)$=0.54,
$(B-V)$=0.96 (Tinsley 1978); Sc or later spiral galaxies with
active star formation typically have colors in the range of 0.35
$\leq$ $(B-V)$ $\leq$ 0.65, -0.2 $\leq$ $(U-B)$ $\leq$ 0.4 (Huchra
1977; Bothun 1982), with mean values $<(B-V)> \sim$0.50, $<(U-B)>
\sim$-0.20 (Bothun 1982), and $<(V-I)> \sim$1.0 (Han 1992).

The colors of our galaxies range continuously from very blue to
very red and include a group of old galaxies, which show evidence
of recent star formation. These LSBGs at the present epoch define
a wide range of evolutionary states. This wide range is similar to
that of the high surface brightness galaxies (see
Sect.~\ref{sec.4}).

\section{Comparisons between the HSBGs and the LSBGs}
\label{sec.4}

\begin{figure}
\centering
\includegraphics[width=6.0cm, height=11.0cm, clip] {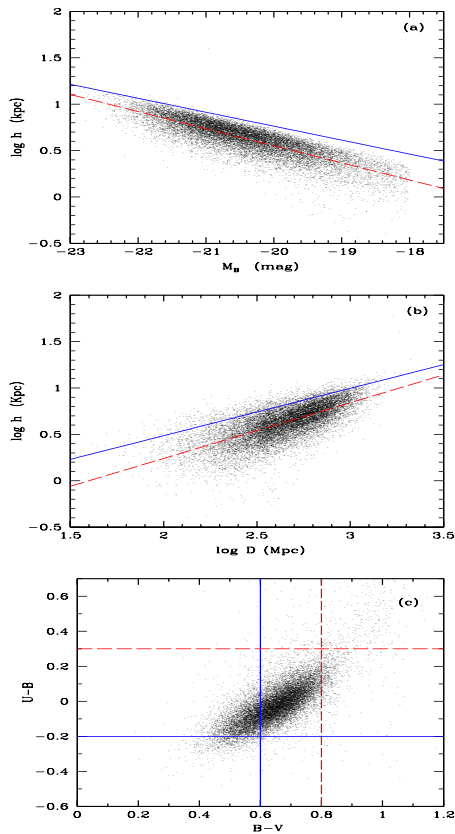}
\caption{ Some relations of HSBGs: (a). correlation of luminosity
and disk scale length, the dashed line is the least-square fit for
the data and the solid line is the fit for the LSBGs given in
Fig.\ref{fig.4}; (b). correlation of distance and disk scale
length, the dashed line is the least-square fit for the data and
the solid line is the fit for the LSBGs given in Fig.\ref{fig.5};
(c). $U-B$ vs. $B-V$ color-color diagram, the solid and dashed
lines are the same as in Fig.\ref{fig.8}.
      }
\label{fig.9}
\end{figure}
%
In this section, we present the correlations between log$h$ and
$M_B$, log$h$ and distance for HSBGs, and compare them with those
of the LSBGs.

Figure~\ref{fig.9}a shows the clear correlation of luminosity and
disk scale length for the HSBGs. This correlation can also be
given by a least-square fit by the dashed line as:

\begin{eqnarray}
\log h = -0.184(\pm 0.0009) M_B - 3.135(\pm 0.018),
\end{eqnarray}
with the standard deviation of 0.10 dex. The Spearman-Rank
Order correlation coefficient is -0.846. The solid line refers to
the corresponding least-square fit for the LSBGs given in
Fig.\ref{fig.4}, which is shallower than the dashed line. The
discrepancy between the solid and dashed lines means that at a
given luminosity, the LSBGs need to be larger than the HSBGs if
both of them can be detected and observed; and with a given disk
scale length, the HSBGs are brighter than the LSBGs.

Figure~\ref{fig.9}b shows the correlation between the distance and
disk scale length for the HSBGs. This correlation can be given by
a least-square fit (the dashed line in the figure) as:
\begin{eqnarray}
 \log h = 0.601(\pm 0.005) \log D - 0.961(\pm 0.012),
\end{eqnarray}
with the standard deviation of 0.14 dex. The Spearman-Rank
Order correlation coefficient is 0.710. The same, the solid line
refers to the least-square fit for the LSBGs. The discrepancy
between the solid and dashed lines means that at a given distance,
the LSBGs need to be larger than the HSBGs if they can be detected
and observed, in another word, with a given disk scale length, the
HSBGs could be observed at a farther distance.

Figure~\ref{fig.9}c shows the color-color diagram of $(U-B)$ and
$(B-V)$ of the HSGBs. The ranges of the two colors are almost
similar to those of the LSBGs, but the HSBGs show less scatter
here. It also shows that the HSBGs have larger fraction of the
``very red" objects than in the LSBGs, i.e., the fraction is
2.1\%, higher than the 0.48\% in the LSBGs, and the fraction of
the ``very blue" objects in the HSBGs is about 4.0\%, which is a
bit lower than the 4.4\% in the LSBGs. This is acceptable since
the HSBGs could have more older stellar populations than the
LSBGs. The others are the mix of old and young stellar
populations.

\section{The volume-limited sub-sample}
\label{sec.55}

Since we are looking for the property relations of low
surface brightness galaxies, it is important to discuss the
completeness of the sample at the low surface brightness end. As
Blanton et al. (2005b) discussed, the incompleteness at low
surface brightness is due primarily to two effects. First, the
inappropriate shredding by the photometric deblender of low
surface brightness galaxies, which often relates to the presence
of nearby stars. This shredding tends to reduce the galaxy fluxes
to well below the flux limits of the survey. Second, for many
galaxies the flux is significantly reduced because the sky
subtraction determination subtracts a substantial fraction of the
galaxy light.

 Blanton et
al. (2005b) discussed the contributions to completeness as a
function of surface brightness, the $r$-band Petrosian half-light
surface brightness $\mu_{50,r}$. In their Fig.3, they present the
completeness of the SDSS photometric catalog, the tilling catalog
(the SDSS spectroscopic targeting with respect to the photometric
catalog), the redshift of the spectroscopy for targets that have
been observed, and the total of these three. It shows that, for
those brighter ones with $\mu_{50,r}<$23 mag arcsec$^{-2}$, the
completeness of the spectroscopy is very close to 100\%, which is
consistent with Strauss et al. (2002), who discussed the high
completeness of the SDSS main galaxy sample. This $\mu_{50,r}$ cut
just corresponds to $\mu_{0}$(B)=24.5 mag arcsec$^{-2}$, which
could be obtained from the fitted relation for our whole sample
galaxies: $\mu_{0}$(B)=1.111$\times$ $\mu_{50,r}$-1.274. But for
the photometric and tilling catalogs, they show obvious
incompleteness within $\mu_{50,r}$=22-23 mag arcsec$^{-2}$
(corresponding to $\mu_{0}$(B) about 23.2-24.3 mag arcsec$^{-2}$),
then the total completeness there could decrease to be about 70\%.

\begin{figure}
\centering
\includegraphics [width=6.0cm, height=11.0cm] {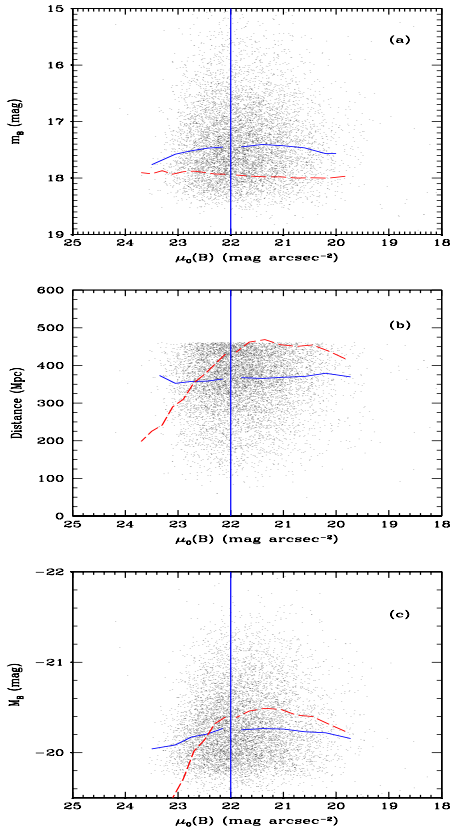}
\caption{Distributions of the property relations of the
volume-limited sub-sample corresponding to Fig.~\ref{fig.3}: (a).
the total $B$-band apparent magnitudes with $\mu_0$(B); (b). the
distance of the galaxies with $\mu_0$(B), the upper cut
corresponds to the redshift cut, $z<$0.1; (c). the $B$-band
absolute magnitudes with $\mu_0$(B). The solid lines are the
median-value points within the bins of $\mu_0(B)$. The
long-dashed lines are the same as Fig.~\ref{fig.3}.}
\label{vol-lim}
\end{figure}

\begin{figure}
\centering
\includegraphics [width=6.5cm, height=11.0cm] {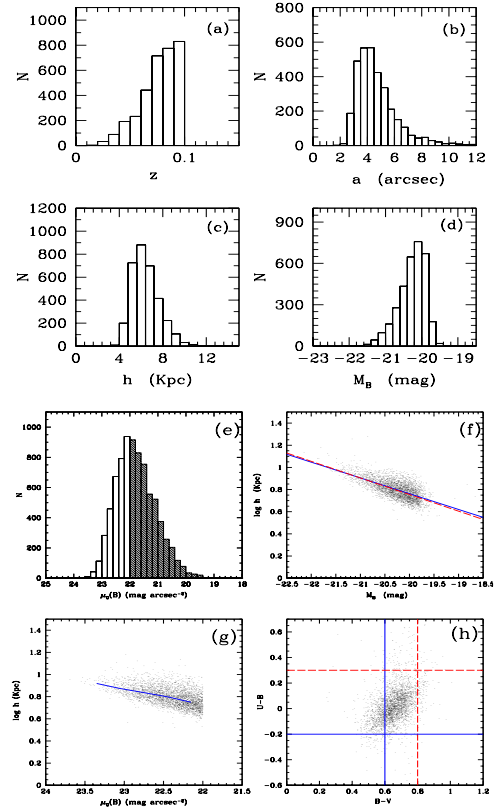}
\caption{ The property parameters of the LSBGs in the
volume-limited sub-sample. (a-d) The same histogram distributions
as Fig.~\ref{fig.2}a-d. (e). Histogram distribution of $\mu_0$(B)
(as Fig.~\ref{fig.1}), the blank histogram is for the 3313 LSBGs
and the shadowed one is for the 4722 HSBGs with the separating
value of 22 mag arcsec$^{-2}$. (f). Disk scale length vs.
luminous, with the solid line as the linear least-square fit, and
the long-dash line is the fit for the entire sample of LSBGs given
in Fig.~\ref{fig.4}. (g). Distribution of disk scale length and
central surface brightness, and the solid line is for the
median-value points (as Fig.~\ref{fig.6}). (h). $UBV$ color-color
diagram (as Fig.~\ref{fig.8}).} 
\label{vol1}
\end{figure}

To avoid the effects of the incompleteness on the property
relations derived from our sample galaxies, and to check that the
obtained trends (Fig.~\ref{fig.1}-\ref{fig.8}) are real or arise
from selection effects, we
 extract a volume-limited sub-sample from the $M_r-z$ plane
 by considering $z<0.1$ and those brighter ones than the corresponding $M_r$,
 and then re-obtain such
relations, which are given by Fig.~\ref{vol-lim} and
Fig.~\ref{vol1}. This volume-limited sample includes 3313 LSBGs
and 4722 HSBGs.

Corresponding to Fig.~\ref{fig.3}, Fig.~\ref{vol-lim} shows the
distributions of this volume-limited sub-sample in the relations
between $m_B$, distance, $M_B$ and $\mu_0$(B). This sub-sample is
similar to that entire sample selected from magnitude in their
relations of $m_B$ and $\mu_0$(B) (Fig.~\ref{vol-lim}a). The upper
cut of distance is the redshift cut of $z=0.1$, and both of the
LSBGs and HSBGs do not show correlations between distance and
$\mu_0$(B) now (Fig.~\ref{vol-lim}b). For the LSBGs in this
volume-limited sub-sample, there exists a weak correlation between
$M_B$ and $\mu_0$(B), but no correlation for the HSBGs
(Fig.~\ref{vol-lim}c). However, because of the redshift cut and
the corresponding $M_r$ magnitude cut for this volume-limited
sub-sample, some distant brighter objects and the nearby faint
ones are lost from the entire sample, thus the range of $M_B$
becomes narrower.

To further check whether the property relations of sample-L
galaxies presented in
Figs.~\ref{fig.1},\ref{fig.2},\ref{fig.4}-\ref{fig.8} are kept or
not in this volume-limited sub-sample, we re-obtained all the related
plots. Fig.~\ref{vol1} shows some of them. Fig.~\ref{vol1}a-d
show the same histogram distributions as Fig.~\ref{fig.2}, but for
the volume-limited sub-sample. They show that the median values of
redshift, disk scale length $a$ and $h$, $B$-band absolute magnitude
 are 0.079, 4.35 arcsec and 6.27 kpc, -20.21 mag,
respectively. Fig.~\ref{vol1}e presents the histogram distribution of
$\mu_0$(B), which is similar to Fig.~\ref{fig.1} for the entire
sample, and with median values of 22.37 for the 3,313 LSBGs and 21.43
for the 4,722 HSBG, srespectively. Fig.~\ref{vol1}f shows the relations
of log$h$ versus $M_B$ (the solid line refers to the linear
least-square fit), which corresponds to Fig.~\ref{fig.4} and is
quite similar to that (the dashed-line). Fig.~\ref{vol1}g shows
that there is a correlation between log$h$ and $\mu_0$(B), which
was not seen in Fig.~\ref{fig.6} for the magnitude-limited sample
galaxies. Fig.~\ref{vol1}h show the $U-B$ vs. $B-V$ color-color
diagram, which is quite similar to the magnitude-limited sample
given in Fig.~\ref{fig.8} and means that most of these LSBGs have
mix of young and old populations. For other relations related to
distance, it is not interesting to show the distribution because
of the redshift (distance) cut. For the color vs. log$h$ and
$\mu_0$(B) relations, this volume-limited sub-sample shows similar
trends to our entire magnitude-limited LSBGs (Fig.~\ref{fig.7}),
and the weak correlation in Fig.~\ref{fig.7}a disappear here. We
did not present them as plots.

\section{Optical and near infrared surface photometry and stellar populations of LSBGs}
\label{sec.5}

The color-color diagram of galaxies can give some hints about
their ages and metallicities and hence the stellar populations.
But the well-known problem in this is the degeneracy of age and
metallicity of galaxies. The recent studies show that the
combination of optical and NIR colors could break this degeneracy.
 Thus several optical-NIR color-color diagrams have been used for
this purpose for the LSB galaxies (Romanishin et al. 1982;
Bergvall et al. 1999; Galaz et al. 2002, 2006; Bell et al. 1999,
2000; Monnier Ragaigne et al. 2003; MacArthur et al. 2004; Bell
$\&$ de Jong 2000 for spiral galaxies). In this section, we obtain
and present the optical-NIR color-color diagram ($(R-K)$ vs.
$(B-R)$ as an example) of our whole sample galaxies and show their
stellar populations by comparing with the stellar population
synthesis models with different metallicities and ages.

\subsection{Cross correlation with 2MASS}

The 2MASS is a ground-based, near-infrared full-resolution for each
extended imaging survey of the whole sky and its Extended Source
Catalog (XSC) contains almost 1.6 million galaxies (Skrutskie et al.
1997; Cutri et al. 2000; Jarrett et al. 2000a,b). 2MASS has imaged
the entire celestial sphere in the near-infrared $J$ (1.11-1.36mm),
$H$ (1.50-1.80) and $K_s$ (2.00-2.32) bands (we use K hereafter)
using two identical dedicated 1.3-meter telescopes for detecting
brighter than 14th mag at $K$ with angular diameters greater than
$\sim$10 arcsec. It has a 95\% completeness level in $J$, $H$ and
$K$ of 15.1, 14.3 and 13.5 mag, respectively.

The NYU-VAGC contains a matched sample between the SDSS and 2MASS
XSC. We do cross-correlation between Sample-L (as well Sample-H)
and the 2MASS-XSC catalog matched by NYU-VAGC with the error
ellipse within 3 arcsec. The resulting samples are { \bf
Sample-L2} with 1,878 (15.29\%) LSB galaxies and { \bf Sample-H2}
with 5,320 (29.47\%) HSB galaxies, respectively. The fraction
detected by the 2MASS NIR for the LSBGs is only half that of the
HSBGs. This can be understood naturally because the galaxies with
lower surface brightness could be lost more easily in a shallower
survey, or due to their relatively bluer colors.

McIntosh et al. (2006) has carefully discussed the completeness of
2MASS. From the cross correlation with the main galaxy sample of
the SDSS, they found that 2MASS detects 90\% of SDSS galaxies
brighter than $r=$17 mag. These detections span the representative
range of optical and near-infrared galaxy properties, but with a
surface brightness-dependent bias that preferentially misses
sources at the extreme blue and low-concentration end of parameter
space, which are consistent with the most morphologically
late-type galaxy population. An XSC completeness of 97.5\% is
achievable at bright magnitudes, with blue low-surface-brightness
galaxies being the only major source of incompleteness. Even with
this bias of surface brightness, our qualitative conclusion about
the mix of stellar populations of LSBGs is still reasonable as
will be discussed in Sect.5.2.

We should guarantee that the SDSS and 2MASS magnitudes are measured
within the same aperture when we obtain the optical-NIR colors of
the galaxies. We adopt the method of Chang et al. (2006a,b) to do
the aperture corrections for the magnitudes, i.e., to correct the
SDSS magnitudes to the same aperture where the 2MASS magnitudes are
measured. To do so, we firstly adopt the isophotal fiducial
magnitudes of $J$, $H$ and $K$ magnitudes provided by the 2MASS,
which are measured within the circular aperture corresponding to a
surface brightness of 20.0 mag arcsec$^{-1}$ in the $K$-band. The
aperture was denoted by $R_{K20fc}$. Then we use the $ProfMean$
measurements provided by the SDSS to match the circular aperture of
the $R_{K20fc}$ in 2MASS. The $ProfMean$ parameter is the
azimuthally averaged surface brightness in a series of 15 circular
annuli, which has been given in all the SDSS $u$,$g$,$r$,$i$,$z$
photometric magnitudes. All the magnitudes are corrected for
Galactic extinction and $K$-correction following the same method
mentioned in Sect.~\ref{sec.2.1}.

\subsection{Optical-NIR color-color diagram and stellar populations}

\begin{figure}
\centering
\includegraphics[width=7.8cm, height=9.0cm, clip]{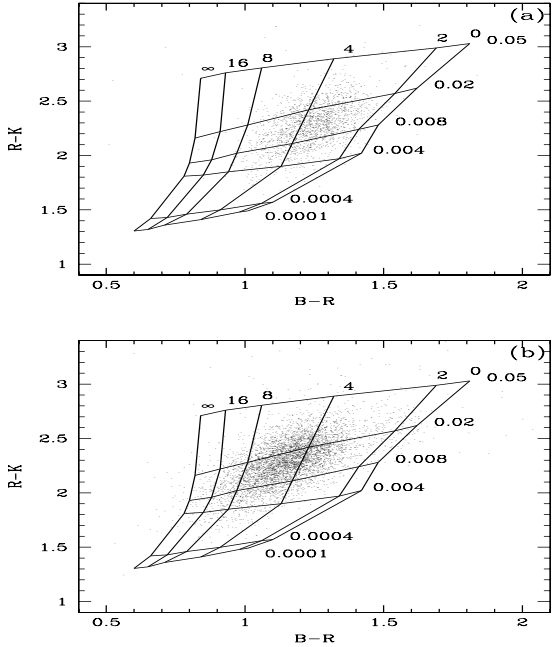}
\caption{Trends in the $B-R$ and $R-K$ optical-NIR color-color
diagram for Sample-L2 in (a), Sample-H2 in (b)). The grids with
different metallicities and ages are given by Bell et al. (2000)
by using the GISSL98 implementation of the stellar population
models of Bruzual $\&$ Charlot (2003). } 
\label{fig.10}
\end{figure}

 In Fig.~\ref{fig.10} we show the optical-NIR  diagrams of
($R-K$) vs. ($B-R$): (a) for Sample-L2; (b) for Sample-H2. The
overplots are the stellar population synthesis models obtained by
Bell et al. (2000) using the GISSL98 implementation of the stellar
population models of Bruzual $\&$ Charlot (2003), where the
horizontal lines refer to the different metallicities and the
vertical lines refer to the different characterized e-folding
time-scale $\tau$ of their star formation. They adopted a Salpeter
(1955) initial mass function (IMF) where the lower mass limit of
the IMF was 0.1 $M_\odot$ and the upper mass limit was 125
$M_\odot$, and an exponentially decreasing star formation rate
characterized by an e-folding time-scale $\tau$ and a single,
fixed stellar metallicity $Z$. Given the joint evidence of the
WMAP that the last electron scatter was about 13.7 billion years
ago and that stars formed 200 million years afterward (Bennett et
al. 2003), it is reasonable to fix the model star formation at 12
Gyrs ago. It was assumed that the IMF does not vary as a function
of time and galactic environment.

\begin{figure}
\centering
\includegraphics[bb=20 167 587 708,width=8cm, height=6.8cm, clip]{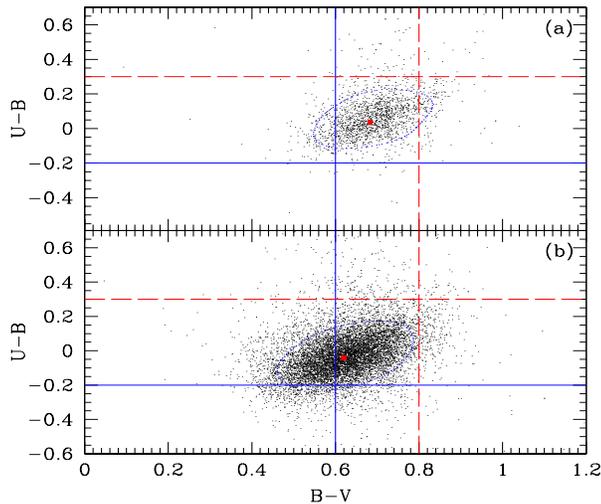}
\caption{ $U-B$ versus $B-V$ diagrams of the 2MASS-detected LSBGs
(fig.a, our Sample-L2) and the non-detected ones (fig.b). The
horizontal and vertical lines as in Fig.~\ref{fig.8}. The contours
show the 68.3\% confidence level of the data points and the large
squares in the center show the peaks of them (please see the
online color version for more details). }  
\label{fig.UBV2mass}
\end{figure}

These results show that the LSBGs have a mix of old and young
stellar populations with metallicities $Z=0.0004-0.05$ and ages,
or the e-folding time scale of star formation rate, $\tau =0-16$
Gyrs. So the continuous range of colors, from very blue to very
red, clearly shows that LSB galaxies at the present epoch define a
wide range of evolutionary states. The HSBGs also show a similar
trend in this optical-NIR color-color diagram although some of
them show a bit redder colors. Although part of the low surface
brightness galaxies are excluded when the 2MASS criteria is
applied (see Sect.5.1), this qualitative conclusion of mixed
population of LSBGs is still believable.
Fig.~\ref{fig.UBV2mass}a,b present the detected (Fig.a) and
non-detected (Fig.b) LSBGs by 2MASS in the relation of $(U-B)$ vs.
$(B-V)$, respectively. The contours show the 68.3\% confidence
level of the data points and the large squares in the center show
their peaks. Although the 2MASS-detected galaxies are biased a bit
to the population with redder colors than the non-detected ones by
2MASS, most of the detected ones (Sample-L2) still show a mix of
young and old stellar populations since they belong to the ``rest
case" of stellar populations suggested by O'Neil et al. (1997a,b).

\subsection{Color-magnitude diagrams}
\label{sec.5.3}

It was known that early-type galaxies in nearby clusters exhibit a
tight color-magnitude relation (CMR), that is, more luminous
early-type galaxies tend to have redder colors. This relation has
been mainly ascribed to the metallicity effect (Chang et al. 2006a
and the references therein).

The  tight CMR of late-type galaxies have also been identified by
many studies, such as Visvanathan $\&$ Griersmith (1977), Blanton
et al. (2003), Baldry et al. (2004), but with a different slope
than the early-type galaxies. The optical-NIR CMR of spiral
galaxies have been obtained for some samples. Chang et al. (2006b)
present the CMR of a large sample of spiral galaxies selected from
the SDSS and 2MASS. They found that the colors for faint galaxies
have a very weak correlation with the luminosity, whereas the
colors for bright galaxies are redder than those for less luminous
galaxies. By comparing with stellar population synthesis model,
they found that massive late-type galaxies have older and
higher-metallicity stellar populations than those of less-massive
galaxies.

\begin{figure}
\centering
\includegraphics [width=7.8cm, height=8.8cm] {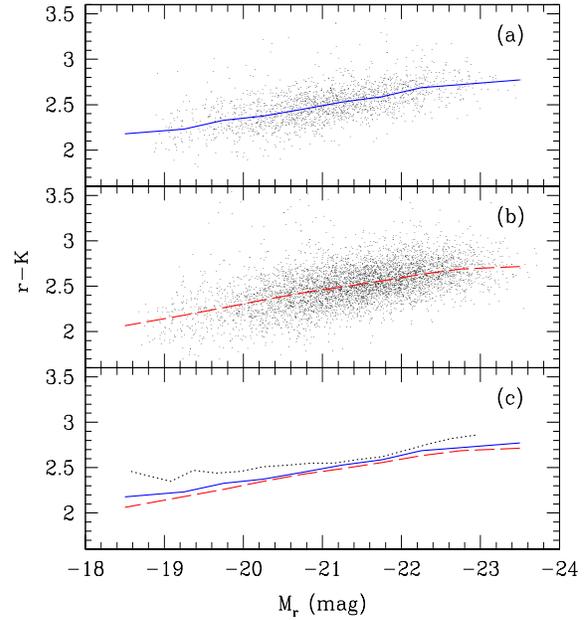}
\caption{Color-magnitude relations of the sample galaxies. The
optical-NIR $r-K$ color and the absolute magnitude in $r$-band
$M_r$ are contained: (a). Our sample of Sample-L2. (b). Our sample
of Sample-H2. (c). It shows the comparisons among Sample-L2, (the
solid line), Sample-H2, (the long-dashed line), and the results of
Chang et al. (2006b, the dotted lines) for spiral galaxies.
   }
\label{fig.11}
\end{figure}

In Fig.~\ref{fig.11}a we present the CMR for Sample-L2 for the
relation of optical-NIR ($r-K$) colors and $r$-band absolute
magnitudes $M_r$. Fig.~\ref{fig.11}b shows the same relations for
Sample-H2. It seems there is no obvious difference between the
LSBGs and the HSBGs in this relation, but the scatter of the HSBGs
are a bit larger than those of the LSBGs. Both of them show that
the brighter galaxies have redder colors than the less bright
ones. Fig.~\ref{fig.11}c shows the comparisons among the results
of Sample-L2, (the solid line), Sample-H2, (the long-dashed line),
and that result of Chang et al. (2006b, the dotted line) for the
SDSS spiral galaxies in the color-magnitude diagram. It shows the
larger sample of spiral galaxies in Chang et al. (2006b) is a bit
redder in $(r-K)$ color than our samples at a given magnitude,
which may be because we selected the galaxies with less
bulge-component with $fracDev_r<$0.25, whereas they adopted
$<$0.5, or these LSBGs had less star formation in the past.
However, maybe it is not very reliable to compare with Chang et
al. (2006b) here since we did not exactly follow them to do the
emission-line corrections for the CMR, because we think it
strongly depends on models. However, the general trend of brighter
galaxies showing redder $(r-K)$  colors is real for these
galaxies. Other colors, such as $(g-r)$, $(r-i)$, $(r-z)$,
$(r-J)$, and $(J-K)$, as presented in Chang et al. (2006b), also
show a similar trend in their relations with $M_r$.

\section{Discussions and conclusions}
\label{sec.6}

Low surface brightness galaxies (LSBGs) are important populations
in the Universe. They were studied little in the past due to the
observational limits of their low surface brightness. The
fantastic modern SDSS survey of large sky coverage area is unique for such studies
because it detects as many low surface brightness galaxies. We
select a large sample of 12,282 LSBGs from the main galaxy sample
of SDSS-DR4 database, and present their basic properties in this
paper. The results are summarized as follows.

\begin{enumerate}

\item Firstly, we select 30,333 nearly face-on disk galaxies with
$M_B<-18$ from the main galaxy sample of SDSS-DR4. Their surface
brightness profiles can be fitted well by an exponential disk. The
$B-$band central surface brightness of these disk galaxies
distribute in a range of 18.5 $<$$\mu_0(B)$$<$ 24.5 mag
arcsec$^{-2}$.
This sample was then divided into two sub-samples: Sample-L
(12,282, the LSBGs) and Sample-H (18,051, the HSBGs) by the
surface brightness value of 22 mag arcsec$^{-2}$. The median
$\mu_0(B)$ of Sample-L is 22.42 mag arcsec$^{-2}$ and the mean
value is 22.50 mag arcsec$^{-2}$. The median and mean values of
$\mu_0(B)$ of Sample-H are 21.41 and 21.29 mag arcsec$^{-2}$,
respectively.

\item The scale length $h$ of Sample-L are from 2 kpc to 19 kpc
with the median scale length of 5.98 kpc. These values are similar
to those scale lengths of Sample-H, which means that the LSBGs and
HSBGs are comparable in size.

\item The absolute $B$-band magnitude $M_B$ of Sample-L are from
-18 to -23 mag with the median value of -20.08 mag and the mean
value of -20.06 mag. These values are similar to those $M_B$ of
Sample-H (-20.47 and -20.41 mag respectively), which means that
the LSBGs are not necessarily faint, and their luminosities are
comparable to those of the HSBGs.

\item Some relations between the property parameters show the
selection effects of the survey observations, such as the apparent
magnitude vs. $\mu_0(B)$, distance versus $\mu_0(B)$,  absolute
magnitude versus $\mu_0(B)$, distance versus disk scale length,
and the distance versus absolute magnitude.
These relations show that the galaxies with higher surface
brightness tend to be detected at farther distance; and the
galaxies observed at farther distance are also brighter and have
larger disk scale lengths.

\item A fundamental correlation to the sample galaxies is the disk
scale length log$h$ versus absolute magnitude $M_B$. For Sample-L,
the scale length log$h$ and absolute magnitude $M_B$ show very
tight correlation, meaning that the brighter galaxies tend to be
larger. The tight correlation between log$h$ and $M_B$ also exists
in the HSGBs, but with a slightly steeper slope, which means that,
at a given $M_B$, the observed LSBGs tend to be larger than the
HSBGs, and for a given size of galaxies, the observed HSBGs tend
to be brighter than the LSBGs.

\item  There is no obvious correlation between colors and central
surface brightness $\mu_0(B)$ for our large sample of LSBGs. The
correlation between colors and disk scale length (log$h$) is weak
even though it exists.

\begin{figure}
\centering
\includegraphics [width=7.8cm, height=9.0cm] {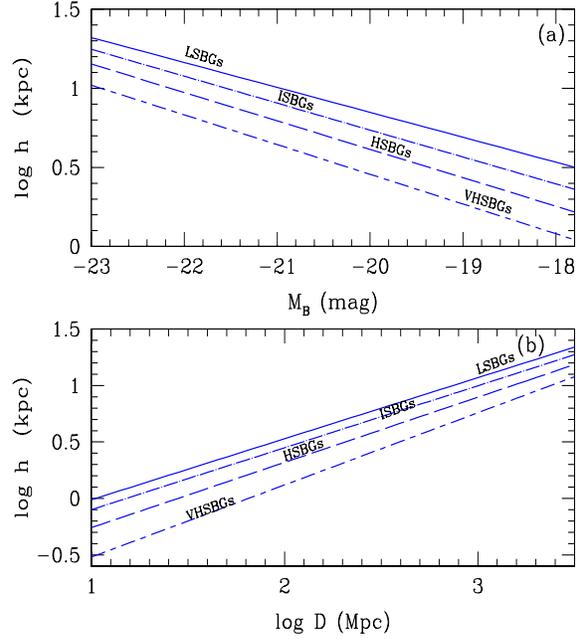}
\caption{(a). Correlation of disk scale length and luminosity for
the whole sample galaxies: LSBGs (the solid line); ISBGs (the
dot-long dash line); HSBGs (the long-dash line); VHSBGs (the
short-long dash line). (b). Correlation of disk scale length and
distance of the whole sample galaxies, and the lines represent the
same means as (a).
      }
\label{fig.14}
\end{figure}

\item The color-color diagrams of $(U-B)$ vs. $(B-V)$ show that
most of our LSBGs have a mix of young and old stellar populations.
Following the definitions of O'Neil et al. (1997a,b), our sample
may contain about 4.4\% (535 out of 12282) ``very blue" LSB
galaxies with $(U-B)$$<$-0.2 and $(B-V)$$<$0.6, and $\sim$0.48\%
(69 out of 12282) ``very red" LSB galaxies with $(U-B)$$>$0.3 and
$(B-V)$$>$0.8. These fractions are 4.0\% and 2.1\% for our HSBGs,
respectively.

\item A volume-limited sub-sample was extracted by considering
$z<0.1$ and brighter than the corresponding $M_r$, which is
complete and can be used to check the incompleteness of surface
brightness (especially at the low end), and to check how the
property relations change. The only one with obvious change is the
relation of log$h$ versus $\mu_0$(B), which shows a correlation in
this sub-sample.

\item The optical-NIR color-color diagrams can break the
degeneracy of age and metallicity. By doing cross-correlations
between the SDSS and 2MASS datasets, we obtain the NIR color
sample of Sample-L2 (15.29\% of the Sample-L), and such sample of
Sample-H2 (29.47\% of the Sample-H). The smaller detected fraction
of LSBGs is not unexpected.

The color-color diagram of $(R-K)$ vs. $(B-R)$ clearly shows the
stellar populations of the sample galaxies by comparing with the
predictions of the stellar population synthesis models. This shows
that our LSBG samples have a wide range of ages and metallicities,
from Z=0.0004 to Z=Z$_{\odot}$, even Z=0.05, and the exponential
infall time-scale $\tau$ is from 0 to 16 Gyrs. It confirms that
most LSBGs have a mix of young and old stellar populations.

\begin{figure}
\centering
\includegraphics [width=7.5cm, height=7.2cm] {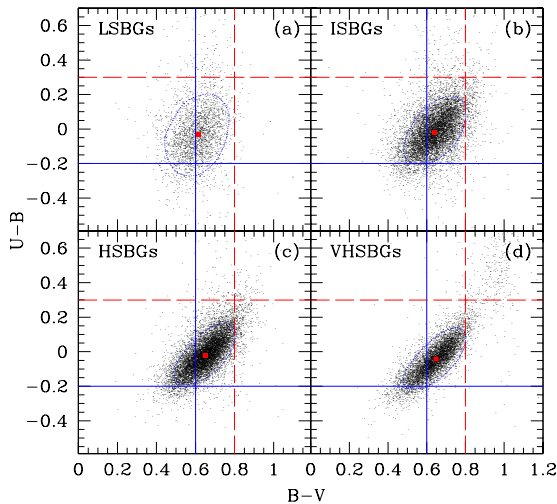}
\caption{ $U-B$ versus $B-V$ diagram: (a). LSBGs, (b). ISBGs, (c).
HSBGs, (d). VHSBGs. The horizontal and vertical lines are the same
as in Fig.~\ref{fig.8}. The contours and the squares are the same
as in Fig.~\ref{fig.UBV2mass} (please see the online color version
for more details). } 
\label{fig.15}
\end{figure}

\item The color-magnitude relations ($(r-K)$ vs. $M_r$) of our
LSBGs confirm that there exist the CMR for the spiral galaxies
with low surface brightness, showing that the brighter LSBGs tend
to be redder. And, the HSBGs do not show obvious differences from
the LSBGs on their CMRs.

\item We have attempted to divide our total nearly face-on disk
galaxies (30,333) into four sub-groups according to their surface
brightness by following McGaugh (1996):
the VHSBGs with $\mu_0(B)$$<$21.25 mag arcsec$^{-2}$,
the HSBGs with 21.25$<$$\mu_0(B)$$<$22 mag arcsec$^{-2}$,
the ISBGs with 22$<$$\mu_0(B)$$<$22.75 mag arcsec$^{-2}$) and
the LSBGs with 22.75$<$$\mu_0(B)$$<$24.5 mag arcsec$^{-2}$.
Similar to the comparisons between LSBGs and HSBGs in
Sect.~\ref{sec.4}, these four sub-groups do show obvious
difference in the relations of log$h$ vs. $M_B$, log$h$ vs.
log$D$, and some difference in the color-color diagram of $(U-B)$
vs. $(B-V)$ as given in Fig.~\ref{fig.14} and Fig.~\ref{fig.15}.
Fig.~\ref{fig.14}a shows that, at a given $M_B$ or scale length
log$h$, the observed galaxies in the four sub-groups show the
gradual increase in their scale length or gradual decrease in
their luminosity following the decreasing surface brightness. The
four lines refer to the linear least-square fits for the data in
the four sub-groups. To show the scatter of the data as well, we
obtained their standard derivations as 0.057, 0.049, 0.050, 0.093
dex for the LSBGs, ISBGs, HSBGs and VHSBGs, respectively.
Fig.~\ref{fig.14}b also shows such gradual changes in their
relations of scale length and distance. Similarly, the four lines
refer to the linear least-square fits for the data in the four
sub-groups, and the standard derivations of the data to the
fittings are 0.092, 0.096, 0.098, 0.13 dex for the LSBGs, ISBGs,
HSBGs and VHSBGs, respectively. Fig.~\ref{fig.15}a-d presents the
discrepancies between their stellar populations among the four
sub-groups. Even with large scatter, the LSBGs do generally show
slightly bluer ($B-V$) color than others, and there are almost no
the ``red populations" in the LSBGs. It also shows the VHSBGs
consist of more ``red populations"  than other groups. The
counters refer to the 68.3\% confidence level of the data points,
which show the gradually redder $(B-V)$ colors of the sub-groups
following their increasing surface brightness.

In summary, this large sample (12,282) of LSBGs greatly extends
the known sample of LSBGs up to date. This is the first paper of
our series work about the large sample of LSBGs to present the
distributions of some property parameters of them, and to provide
more information about some of their fundamental properties, e.g.
the tight correlations between log$h$ and $M_B$, log$h$ and
log$D$, moreover, it presents their stellar populations by using
the color-color diagrams of them.

In the following work, we will discuss more about the major issues
of these ``populations", such as why they are LSBGs with small
star formation rates, what is their total mass content and M/L
ratios, and what is their total contribution to the baryonic/total
mass etc. We will also obtain their metallicities in interstellar
gas from emission lines. The detailed surface photometry and color
gradients will be also studied in detail. Moreover, some more
efforts on other aspects are also needed, such as to extend the
studies of edge-on galaxies, but the dust effects must be
considered carefully then.

\end{enumerate}

\section*{Acknowledgments}

 We thank our referee for the valuable comments and
suggestions, which help to improve well our work. We thank
Ruixiang Chang, Zhengyi Shao, Jianling Wang for helpful
discussions, and thank James Wicker for English corrections on the
text. We thank the NSFC grant support under Nos. 10403006,
10433010, 10673002, 10573022, 10333060, and 10521001, and the
National Basic Research Program of China (973 Program)
No.2007CB815404, 06 and the Doctoral Foundation of SYNU of China
(054-55440105020).

\label{lastpage}

\end{document}